# Lightweight Efficient Multi-keyword Ranked Search over Encrypted Cloud Data using Dual Word Embeddings

Ruihui Zhao[†]   Mizuho Iwaihara[‡]

Graduate School of Information, Production, and Systems, Waseda University

2-7 Hibikino, Wakamatu-ward, Kitakyushu-city, Fukuoka-ken, 808-0135 Japan

E-mail:  [†]zachary@ruri.waseda.jp,  [‡]iwaihara@waseda.jp

**Abstract.** Cloud computing is emerging as a revolutionary computing paradigm which provides a flexible and economic strategy for data management and resource sharing. Security and privacy become major concerns in the cloud scenario, for which Searchable Encryption (SE) technology is proposed to support efficient retrieval of encrypted data. However, the absence of lightweight ranked search is still a typical shortage in existing SE schemes. In this paper, we propose a Lightweight Efficient Multi-keyword Ranked Search over Encrypted Cloud Data using Dual Word Embeddings (LRSE) scheme that supports top-k retrieval in the known background model. For the first time, we formulate the privacy issue and design goals for lightweight ranked search in SE. We employ word embedding trained on the whole English Wikipedia using word2vec to replace the general dictionary, afterwards we make use of Dual Embedding Space Model (DESM) to substitute traditional Vector Space Model (VSM), based on which we achieve the goal of lightweight ranked search with higher precision and solve the challenging problems caused by updating the traditional dictionary in existing SE schemes. In LRSE, we employ an improved secure kNN scheme to guarantee sufficient privacy protection. Our security analysis shows that LRSE satisfies our formulated privacy requirements and extensive experiments performed on real-world datasets demonstrate that LRSE indeed accords with our proposed design goals.

**Keywords:** Searchable encryption, Lightweight, Dual Word Embeddings

## 1    Introduction

Cloud computing is a revolutionary computing paradigm which provides a flexible and economic strategy for data management and resource sharing [1], [2], thus is getting more and more attention from both academic and industry communities. However, security and privacy become major concerns in the cloud scenario when data owners outsource their private data onto public cloud servers to be accessed by the authenticated users. Usually, the cloud server is considered as curious and untrusted entities [3], thus there are risks of data exposure to a third party or even the cloud service provider itself.

Therefore, providing sufficient security and privacy protections on sensitive data is extremely important, especially for those applications dealing with health, financial and government data. To avoid information leakage, the sensitive data has to be encrypted before uploading onto the cloud servers, which makes it a big challenge to support efficient keyword-based queries and rank the matching results on the encrypted data.

In the branch of plaintext information retrieval (IR) and document filtering, such as a common practice in web search engines (e.g., Google search), data users may tend to provide a set of keywords as the indicator of their search interest to retrieve the most relevant data. "Coordinate matching", i.e., as many matches as possible, and ranking matching documents by certain criteria, has been widely used in the plaintext information retrieval (IR) field. In the field of document retrieval and natural language processing, deep learning-based methods, such as word embeddings [4], are emerging as replacing traditional term vectors for measuring relatedness between terms. However, existing techniques in plaintext information retrieval and document filtering cannot be used in SE scenario directly. How to apply existing schemes in the plaintext field to encrypted cloud data search systems remains a challenging task.

To address the issue, searchable encryption (SE) technology has been proposed in the literature in pursuit of search over encrypted data. For schemes [3,5] that realize flexible search, they only support Boolean keyword search or single keyword search and return inaccurate results that are often loosely related to the user's intent. In 2014, Cao et al. [6] firstly proposed an effective mechanism called MRSE to partially solve the multi-keyword ranked search problem according to the number of matching keywords between the query and documents, which established the foundation and basic framework of multi-keyword ranked search in the field of searchable encryption. As far as we know, most of latest schemes in SE follow this framework, such as [7, 8, 9]. As a consequence, they have the same congenital drawbacks. Here we conclude the congenital drawbacks of existing searchable encryption schemes as follows:

1) Low precision. Most of latest existing SE schemes which follow the classic MRSE scheme [6] are based on keyword match-based method, which is functionally inferior in the view of current plaintext information retrieval and machine learning. For example, MRSE simply counts the number of matching keywords between query and documents and does not take the access frequencies of the keywords into account. Although following work such as [6, 8, 9] emploies *tf-idf* weight to substitute occurrence bits in binary vectors, they are still too primal and basic schemes, because *tf-idf* weighting is quite a classic technique in information retrieval field. Besides, in MRSE_II, in order to guarantee security and avoid *Scale Analysis Attack* [6], the authors insert $U$ dummy keywords in subindex $I_i$, which makes the final score deviate from the real score. For example, we can set $U = 200$, and the dimension of the keyword dictionary becomes *4000*. In the framework, they cannot resolve difficulties in achieving both high precision and security, i.e., if they set the standard deviation $\sigma$ of the random variable $\varepsilon$ larger, the final results are totally irrelevant to the query because of the $U$ inserted dummy keywords. Or, if they set $\sigma$ lower, this framework cannot withstand the *Scale Analysis Attack* [6]. In conclusion, these two reasons cause the problem of low precision.

2)   Dimension disaster. In MRSE [6], the system overhead during the whole process of index construction, trapdoor generation, and executing query, is mostly determined by matrix multiplication, in another word, the large dimension of the sparse vector according to the dictionary in this framework causes dimension disaster. For example, in MRSE_II time cost of building index for a data owner is almost *6,700*s, when the number of documents in the dataset is *10,000* and the dictionary has *4,000* keywords, which is too time consuming. This problem leads to a result that searchable encryption schemes cannot be practically put into use in a real-world scenario.

3)   Problems caused by dictionary updates. Because all the dimensions of secret key *SK*, index and trapdoor are determined by the dictionary, an update in the dictionary will lead to a result that all generated index and trapdoor cannot be used and all previous work should be executed again.

4)   Lack of fuzzy search and intelligent search. For example, if data user inputs "Java." in the query, if the corresponding keyword in the dictionary is "java", there is no possibility to return relevant results due to the lack of "Java." in the dictionary and keyword match-based method. Let alone some more advanced examples in existing plaintext information retrieval, such as intelligent search using the semantic similarity, for example, "java" and "python", or "iPhone" and "cellphone". Because keyword match-based method cannot measure semantic similarities in such cases.

In conclusion, further researches are necessary to achieve lightweight efficient multi-keyword ranked search over encrypted cloud data with privacy preserved and higher precision. In this paper, we discuss a deep learning-based approach which supports fuzzy search and intelligent search, and consider useful techniques for updating dictionaries, which is still a challenging problem.

In this paper, we propose a Lightweight Efficient Multi-keyword Ranked Search over Encrypted Cloud Data using Dual Word Embeddings (LRSE) scheme that supports top-k retrieval in the known background model. In summary, this paper makes the following **contributions:**

1)   It firstly establishes a set of design goals and privacy issues for lightweight multi-keyword ranked search in a known background model. The improved kNN scheme guarantees high privacy protection.

2)   For the first time, it introduces deep learning-based method in SE framework. It combines word embedding trained using word2vec and dual embedding space model (DESM), to achieve the goal of lightweight efficient multi-keyword ranked search over encrypted cloud data, providing search results with higher precision and more consistent with the query. Besides, it can also support fuzzy and intelligent search.

3)   It lightens the problems caused by dictionary updates and reduces the system cost even when the vocabulary in our scheme has to be updated, through dimension reduction by utilizing word embeddings.

4)   Thorough analysis investigating privacy and performance evaluation is given, and experiments on the real-world dataset further show the proposed scheme indeed introduces low overhead.

The remainder of this paper is organized as follows: we discuss existing related work on searchable encryption and document filtering based on deep learning in Section 2. In Section 3, we introduce the system model, security requirements, design goals, and preliminary on word embeddings. Section 4 describes the LRSE framework and proposed schemes, followed by Section 5, which focuses on security analysis. Section 6 presents performance evaluation and simulation results. At last we conclude the paper in Section 7.

## 2  Related Work

### 2.1  Existing Searchable Encryption

Cao et al. [6] propose an effective mechanism which can partially solve the multi-keyword ranked query problem according to the number of matching keywords, however, MRSE does not take the access frequencies of the keywords into account. It only returns the documents ordered by the number of matched keywords. Besides, MRSE has problems such as dimension disaster and low precision. Yu et al. [11] propose a two-round searchable encryption that supports top-k multi-keyword retrieval, which can guarantee high security and practical efficiency. But it is also based on keyword-match based method, and has the problems of low precision and lack of intelligent search.

### 2.2  Document Filtering based on Deep Learning

Nalisnick et al. [10] propose a dual embedding space model, which can be used to calculate the similarity for a document and a query term, complementing the traditional term frequency based approach. However, it is not directly applicable in the context of encrypted cloud data retrieval.

## 3  Problem Formulation

### 3.1  System Model

As illustrated in Fig. 1, our scheme involves three different entities.

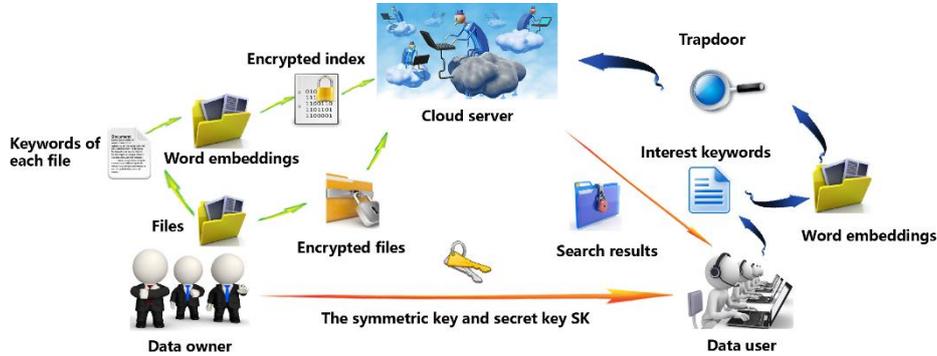

**Fig. 1.** System Model

1. **The cloud server:** The cloud server is an intermediate entity hosting third-party data storage and retrieve services to authenticated data users. When received a trapdoor from the data user, the cloud server will locate the matching files by scanning the indexes *I*, calculate corresponding relevance scores, and return the ranked top-k results to the data user.
2. **The data owner:** The data owner encrypts a collection of files using symmetric encryption algorithm and builds a searchable index *I*, then he outsources both the encrypted indexes *I* and encrypted files *C* onto the cloud server. After that, the data owner sends the symmetric key and secret key *SK* to the data user.
3. **The data user:** The data user generates a trapdoor with *SK* and sends it to the cloud server. Afterward, the data user is returned the most relevant top-k encrypted files and indexes, he decrypts and makes use of them with the help of the symmetric key and *SK*.

Note here, there is a special case where the data owner is the same as the data user, that's to say, the data owner keeps his secret key *SK* by himself and only searches over his own encrypted documents. Then there are only two entities left in this special case: the data owner and the cloud server. This case also has many vivid applications in daily life, for example, when we upload our local documents in the phone and PC to the cloud server, such as: Google, Amazon, Microsoft, and Baidu cloud drive, there is a horrible security problem: all of our passwords, patents to be published, private photos, health records are exposed to the mentioned companies. Then this special case could help solve this problem.

### 3.2 Security Requirements

The cloud server is considered as honest but curious, i.e., it is designed to execute the service algorithm faithfully, however, it is also curious and eager to attain sensitive information. We define the security requirements as follows:

1. Data, index, and trapdoor privacy: Data privacy means that LRSE should prevent the cloud server from poking its nose into the outsourced data. Index privacy means that the index should be constructed to prevent the cloud server from performing association attack i.e., deducing any association between keywords and encrypted documents. Besides, trapdoor privacy means what the user submits according to his interest is well protected by the complexity of trapdoor generating algorithm.
2. Unlinkability of trapdoors: In LRSE, we define the unlinkability of trapdoors in a harsh model, Known Background Model [6]. In this case, the cloud server is more powerful and possesses some statistical information to carry out *Scale Analysis Attack* [6]. We should assure that the cloud server would not be able to identify the keywords in a query even if some background information had been leaked.

### 3.3 Design Goals

In order to realize our LRSE scheme, the following security and performance guarantees should be simultaneously achieved:

1. **Lightweight efficient multi-keyword ranked search:** It can support efficient multi-keyword based query with low overheads and higher precision, guarantee the most relevant files to appear in the top-k locations, and solve the dimension disaster.
2. **Privacy-preserving:** To meet all of the harsh security requirements specified in Section 3.2.
3. **Solving the problem caused by dictionary updates:** As specified in Section 1, traditional MRSE-based schemes utilize a dictionary containing around *10,000* keywords, but which is far from enough. Once the dictionary is updated, all generated indices and trapdoors cannot be used and the whole process should be executed again. We should solve this problem.
4. **Fuzzy and intelligent search supporting:** Supporting fuzzy and intelligent search in searchable encryption is necessary. It can realize various functions in SE as intelligent as today's web search engines (e.g., Google search) and output the data files consistent with the user's interest.

### 3.4 Preliminary on Word Embeddings

Word embeddings [4] are a generic name of a set of NLP techniques, where each unique word is represented by a relatively low dimension vector of real numbers. Their models are learned through two-layer neural networks, to capture linguistic contexts of words. We trained word embeddings using word2vec [4] in this paper. Word2vec helps to represent each word $w$ of the training set as a vector of features, where this vector is supposed to capture the contexts in which $w$ appears.

In this paper, we chose the whole English Wikipedia corpus, containing *2,126,359* words, as the training set. Our training parameters of word2vec are set as follows: continuous bag of words (CBOW) model instead of the skip-gram (word2vec options: cbow = *1*); the output vectors size is set to *100* (word2vec options: size = *100*); the number of negative samples is set to *25* (word2vec options: negative = *25*).

## 4 The Design of LRSE

In this section, we first propose a basic idea for the LRSE by elegantly combining word embedding and dual embedding space model (DESM) with an improved kNN scheme, which mainly consists of the following four phases: GenKey, BuildIndex, GenTrapdoor, and Query.

### 4.1 GenKey

The data owner randomly generates a (*n+2*)-dimension binary vector as *S* and two (*n+2*)×(*n+2*) invertible matrices {*M₁*, *M₂*}, where *n* equals to the dimension of word embeddings we obtained in Section 3.4. We extend the dimension of vectors in our schemes to (*n+2*)-dimension in order to introduce random numbers in the final results to protect the search results. The secret key *SK* is in the form of a *3*-tuple as {*S, M₁, M₂*}.

Note here, *S* is part of secret key *SK*, which is exactly a binary vector, acting as an indicating vector in the spilt process of building index and generating trapdoor. *S* is randomly generated by the data owner using existing Random Number Generation Algorithm (PRGA) in the field of information security, which is not within the scope of this paper.

### 4.2 BuildIndex

The data owner utilizes the tf-idf statistical method to extract keywords, usually top-25 keywords for each document in our scheme. $D_i$ is the keyword set of *i-th* document. $|Di|$ denotes the number of keywords in the *i-th* document. $d_{ij}$ is the embedding vector for the *j-th* keyword of the *i-th* document. Let $\overline{Di}$ denotes the centroid of all the normalized document word vectors serving as a single embedding for document $D_i$. So we define vector $\overrightarrow{D_i}$ using the following equation:

$$\overrightarrow{D_i} = \frac{\overline{Di}}{\|\overline{Di}\|} \qquad (1)$$

*where,*

$$\overline{Di} = \frac{1}{|Di|} \sum_{d_{ij} \in D_i} \frac{d_{ij}}{\|d_{ij}\|} \qquad (2)$$

Then we extend the dimension of $\overrightarrow{D_i}$ from *n* to *n+2*, which means adding a random number $\varepsilon_i$ in the (*n+1*)-dimension and *1* in the (*n+2*)-th dimension. $\overrightarrow{D_i}$ is therefore extended to $(\overrightarrow{D_i}, \varepsilon_i, 1)$. $\varepsilon_i$ obeys the normal distribution, whose standard deviation is $\sigma$, actually $\varepsilon_i$ is used to obscure the query results to resist frequency analysis attack and its range is determined by the range of query results.

Every plaintext subindex $\vec{D_i}$ is then spilt into a data vector pair donated as $\{\vec{D_{i'}}, \vec{D_{i''}}\}$ using the splitting process of the secure *k*-nearest neighbor (kNN) scheme [12] as follows: For *m=1* to *n+2*, if $\vec{S}[m] = 1$, then $\vec{D_{i'}}[m]$ and $\vec{D_{i''}}[m]$ are set to two random numbers so that their sum is equal to $\vec{D_i}[m]$; else, $\vec{D_{i'}}[m]$ and $\vec{D_{i''}}[m]$ are set as the same as $\vec{D_i}[m]$. Finally, the subindex $I_i = \{M_1^T \vec{D_{i'}}, M_2^T \vec{D_{i''}}\}$ is built for every encrypted document $C_i$.

### 4.3 GenTrapdoor

The data user inputs a set of query keywords according to his interest. We define $Q$ is the keyword set of query, thus $|Q|$ is the number of the keywords in the query. Vector $q_k$ is the embedding vector for the *k*-th keyword of the query. Thus $\vec{Q}$ is generated as follows:

$$\vec{Q} = \frac{1}{|Q|} \sum_{q_k \in Q} \frac{q_k}{\|q_k\|} \qquad (3)$$

The vector $\vec{Q}$ is extended to *(n+1)*-dimension, where the *(n+1)*-dimension is set to *1*, then multiplied by a random number $r \neq 0$, and finally extended to a *(n+2)*-dimension vector where the last dimension is set to another random number *t*. $\vec{Q}$ is therefore equal to $(r\vec{Q}, r, t)$. Here only the data user knows the exact values of *r* and *t*.

For *m=1* to *(n+2)*, if $\vec{S}[m] = 0$, then $\vec{Q'}[m]$ and $\vec{Q''}[m]$ are set to two random numbers so that their sum is equal to $\vec{Q}[m]$; else, $\vec{Q'}[m]$ and $\vec{Q''}[m]$ are set the same as $\vec{Q}[m]$. Finally, the trapdoor *T* is generated as $\{M_1^{-1}\vec{Q'}, M_2^{-1}\vec{Q''}\}$ for the query.

### 4.4 Query

For each document $D_i$, with the trapdoor *T*, the cloud server computes the similarity scores as shown in the following equation, ranks all scores and returns the top-*k* ranked files and their corresponding indexes to the data user.

Note that we are not simply replacing the dictionary using word embedding. Because vectors generated from word embeddings cannot be used in the secure kNN framework directly. We utilize the dual embedding space model (*DESM*) model [10] to replace the *VSM* model in *MRSE*-based schemes. Finally we modified the original scheme and achieve our goal of combining DESM model and the secure kNN framewok using Equ.(4-5).

$$I_i \cdot T = \{M_1^T \vec{D_{i'}}, M_2^T \vec{D_{i''}}\} \cdot \{M_1^{-1} \vec{Q'}, M_2^{-1} \vec{Q''}\}$$
$$= \vec{D_{i'}} \cdot \vec{Q'} + \vec{D_{i''}} \cdot \vec{Q''} = \vec{D_i} \cdot \vec{Q}$$
$$= (\vec{D_i}, \varepsilon_i, 1) \cdot (r\vec{Q}, r, t)$$

$$= r\{DESM(D_i,Q) + \varepsilon_i\} + t \qquad (4)$$

where,

$$DESM(D_i,Q) = \frac{\overline{D_i}}{\|\overline{D_i}\|} \cdot \frac{1}{|Q|} \sum_{q_k \in Q} \frac{q_k}{\|q_k\|} \qquad (5)$$

## 5  Security Analysis

In this section, we analyze the security properties under the schemes we introduced above. We will focus on three aspects: data privacy, index and trapdoor privacy, and trapdoor unlinkability.

### 5.1  Data Privacy

Traditional symmetric key encryption techniques (e.g., AES [12]) could be properly utilized here to guarantee data privacy and is not within the scope of this paper.

### 5.2  Index and Trapdoor Privacy

According to the secure kNN scheme [13], if the secret key SK is kept confidential, the index privacy is well protected by the computation complexity of deducing the meaning of every item in the index. In addition, the introduction of random number $\varepsilon_i$ in Build-Index procedure also adds to the complexity and nondeterminicity of the index generation. As a result, even the generated indexes for the same search document $F_i$ by the same data owner at different index building times are always irrelevant because of the randomness in the spilt process [13] and the nondeterminicity of random number $\varepsilon_i$.

Based on the same principle, the produced trapdoors will be irrelevant even to the same query at different trapdoor generating times. Besides, the introduction of random numbers $r$ and $t$ also contributes to the trapdoor privacy. Note here, random numbers $r$ and $t$ is not only used to enhance the trapdoor privacy, but also to make the final query results different even for the same query at different query times.

### 5.3  Trapdoor Unlinkability

The trapdoor should be constructed to prevent the cloud server from deducing the relationships of any given trapdoors and the corresponding keywords, for example, *Scale Analysis Attack* in a *Known Background Model* introduced in [6]. Basically, *Scale Analysis Attack* in *MRSE* schemes is based on sparse binary vectors, which cannot be applied to our scheme because all the initial vectors in our schemes are low-dimensional vectors of real numbers generated from our word embeddings. Thus we can draw a conclusion that we achieve the same security level of trapdoor unlinkability as *MRSE_II* scheme [10] from the point of resisting *Scale Analysis Attack*. Besides, even under the extreme condition that the cloud server decrypts the vector, it still needs as many as $(C_N^1 + C_N^2 + C_N^3 + \cdots + C_N^k)/2$ computation of Equ.3 in average to confirm the keywords concealed behind the word embedding vector, where *k* ranges from *5* to *50*, and *N = 212,6359*. In another word, word embeddings introduced in LRSE help enhance trapdoor unlinkability.

We note, however, that, we cannot protect against Access Pattern [6], which is defined as the sequence of ranked search results. Although our proposed scheme is not designed to protect against Access Pattern, because of the efficiency consideration just as most existing SE schemes (excluding costly PIR technique [7]). To deal with the situation, we add the random number $\varepsilon_i$ in the final result. The precision of our scheme is affected by the standard deviation $\sigma$ of the random number $\varepsilon_i$.

From the consideration of precision, $\sigma$ is expected be smaller to obtain high precision. For example, in an extreme case, $\sigma$ is set to 0, then all the final ranked results are absolutely the same as the true ranked query results, while Access Pattern are totally without protection. Thus the standard deviation $\sigma$ can act as a flexible trade-off parameter to adjust precision and security of Access Pattern.

## 6 Performance Evaluation

### 6.1 Functionality

As illustrated in [10], DESM model performs better than traditional keyword match-based methods, for example, BM25 [10]. While the MRSE schemes are actually based on simply counting the matched keywords number among the index and trapdoor, which is an quite original keyword match-based method. Let alone MRSE_II scheme adds U dummy keywords to the final results, for example, *U = 200*. We can draw a conclusion that our precision is much larger than MRSE-based schemes. Besides, as presented in Section 5.3: the precision can be affected by the standard deviation $\sigma$ of the random variable $\varepsilon_i$. It is obvious that a smaller $\sigma$ leads to better precision. If we do not attempt to protect Access Pattern at all, $\sigma$ could be set as *0*, and it will not affect the true ranked query results.

Our scheme can also support fuzzy search and intelligent search, for example, if we search "Java.", the top-k keywords are as follows: "Java", "swing", "android", "c++", "c#", "python", "eclipse", etc. And if we input "android" as a query keyword, the top-k relevant keywords are: "Android", "iPhone", "blackberry", "emulator", "java", etc.

LRSE also lightens the problems caused by dictionary updates: since the vocabulary size of our word embeddings is large enough, we almost do not need to change the vocabulary. Besides, LRSE can also reduces the system cost even when the vocabulary in our scheme has to be updated, through dimension reduction by utilizing word embeddings.

## 6.2 Efficiency

We conduct a thorough experiment on a real-world dataset: the *NSF research award dataset*[1] and evaluate the performance of LRSE compared with the MRSE schemes [6].

1. ***Building Index:*** The time cost of building the whole index is related to the number of documents and the computation complex of building each subindex. As presented in Section 4.2, the major computation of building a subindex includes four parts: search in word embeddings for each keyword of the document, generate the initial vector using Equ.1, the splitting process and two multiplications of an $(n + 2) \times (n + 2)$ matrix and an $(n + 2)$-dimension vector. As illustrated in Tab.1, given the same size of dataset with *1000* documents, the index construction time of *LRSE* is much less than *MRSE* schemes due to the difference of their vector dimensions, for example, when the dimension of vectors in *LRSE* and *MRSE* are *100* and *4,000* respectively, LRSE is much more lightweight. Tab.2 shows the time cost of building index with different document number when the vector dimension in LRSE is 300, while the keyword number in the dictionary is 4000 in MRSE, we can see the time cost of LRSE scheme is much less. The time cost of building the whole index is almost linear with the size of dataset since the time cost of building each subindex is fixed. Besides, as shown in Tab.3, we compare the storage overhead of subindex in *LRSE*, *MRSE_I*, and *MRSE_II* within different sizes of dictionary. The size of subindex is absolutely linear with the size of dictionary.

**Table 1.** Time cost of building index with different vector dimension

| **Dim_*MRSE*** | 2000 | 4000 | 6000 | 8000 |
|---|---|---|---|---|
| *MRSE_I* (*s*) | 83.676 | 582.053 | 1874.998 | 4319.606 |
| *MRSE_II* (*s*) | 106.489 | 671.911 | 1979.112 | 4608.017 |
| **Dim_*LRSE*** | 50 | 100 | 200 | 300 |
| *LRSE* (*s*) | **5.938** | **6.536** | **9.577** | **13.612** |

**Table 2.** Time cost of building index with different document number

| **Doc_num** | 2000 | 4000 | 6000 | 8000 | 10000 |
|---|---|---|---|---|---|
| *MRSE_I* (*s*) | 1117.463 | 2240.320 | 3404.328 | 4633.588 | 5856.057 |
| *MRSE_II* (*s*) | 1348.232 | 2679.574 | 4026.798 | 5460.893 | 6802.423 |
| *LRSE* (*s*) | **27.124** | **54.787** | **81.986** | **110.642** | **136.600** |

**Table 3.** Size of subindex/trapdoor

| **Dim_*MRSE*** | 4000 | 6000 | 8000 | 10000 |
|---|---|---|---|---|
| *MRSE_I* (*KB*) | 31.2656 | 46.8906 | 62.5156 | 78.1406 |

---

[1] https://kdd.ics.uci.edu/databases/nsfabs/nsfawards.html

| MRSE_II (KB) | 32.8203 | 48.4453 | 64.0703 | 79.6953 |
|---|---|---|---|---|
| **Dim_LRSE** | 50 | 100 | 200 | 300 |
| LRSE (KB) | **0.4063** | **0.76969** | **1.5781** | **2.3594** |

2. *Generating Trapdoor:* Time cost of generating each trapdoor is determined by four parts: search over word embeddings for each keyword in query using HashMap, generate the initial vector using Equ.3, the complexity of the splitting process and multiplications of a matrix and two spilt query vector. As shown in Tab.4, the time of generating a trapdoor is greatly affected by the dimension of vectors. Thus the trapdoor generating time of LRSE is much less than MRSE schemes due to the difference of their dimensions. Besides, as illustrated in Tab.5, the number of keywords in the query has little influence upon the result because the dimension of vector and matrices is always fixed with the same word embeddings, here the vector dimension in LRSE is set to *100*, while the keyword number in the dictionary of MRSE is set to *4,000*. With respect to the size of the trapdoor, it occupies the same space overhead as that of each subindex listed in Tab. 3, which is only determined by the dimension of the word embeddings.

3. *Executing Query:* The major computation to execute a query in the cloud server consists of computing the similarity scores of each index and trapdoor, and ranking similarity scores for all documents in the dataset and selecting top-*k* results from all the scored documents. Thus the executing query time of LRSE is much less than MRSE schemes due to the difference of their dimensions, as shown in Tab.6. we set *k* to *50* in our experiment. We can learn that the query time is linear with both the number of documents in the dataset and the size of the dictionary. In addition, our proposed scheme introduces nearly constant overhead as increasing the number of query keywords for the same reason as described in generating trapdoor.

**Table 4.** Time cost of generating trapdoor with different vector dimension

| **Dim_MRSE** | 2000 | 4000 | 6000 | 8000 |
|---|---|---|---|---|
| MRSE_I (s) | 0.185 | 0.710 | 1.633 | 2.899 |
| MRSE_II (s) | 0.213 | 0.767 | 1.716 | 2.974 |
| **Dim_LRSE** | 50 | 100 | 200 | 300 |
| LRSE (s) | **0.002** | **0.002** | **0.003** | **0.006** |

**Table 5.** Time cost of generating trapdoor with different keyword number in query

| **Keyword num** | 10 | 25 | 40 | 50 |
|---|---|---|---|---|
| MRSE_I (s) | 0.715 | 0.710 | 0.701 | 0.717 |
| MRSE_II (s) | 0.772 | 0.767 | 0.761 | 0.761 |
| *LRSE* (*s*) | **0.002** | **0.002** | **0.002** | **0.002** |

Table 6. Time cost of query with 1000 documents

| **Dim_MRSE** | 2000 | 4000 | 6000 | 8000 |
|---|---|---|---|---|
| *MRSE_I* (*s*) | 0.008230 | 0.015790 | 0.023460 | 0.032200 |
| *MRSE_II* (*s*) | 0.008525 | 0.016324 | 0.026690 | 0.034324 |
| **Dim_LRSE** | 50 | 100 | 200 | 300 |
| *LRSE* (*s*) | **0.000370** | **0.000550** | **0.000980** | **0.001480** |

# 7 Conclusion

In this paper, we propose a Lightweight Efficient Multi-keyword Ranked Search over Encrypted Cloud Data using Dual Word Embeddings (LRSE) scheme, which supports top-k retrieval in the known background model. We first formulate stringent privacy requirements and design goals in LRSE. For the first time, we introduce deep learning-based method in SE framework. We combine word embeddings trained using word2vec and dual embedding space model (DESM), which provides search results with higher precision and more consistent with the query. Besides, we show examples that our scheme can also support fuzzy and intelligent search. Security analysis shows that the proposed scheme accords with our formulated privacy requirements. Extensive experiments based on real-word dataset demonstrate that the proposed scheme achieves the design goals. In the future, we will introduce personalized search and multi-grained security into our scheme. We will also conduct experiments on the famous *Cranfield*[2] *collection* to further verify our precision in the future.

---

[2] http://ir.dcs.gla.ac.uk/resources/test_collections/cran/